\documentclass[letter,11pt]{article}
\usepackage[usenames,dvipsnames]{xcolor}
\usepackage{fullpage}
\usepackage{natbib}
\usepackage{multicol}
\usepackage{localcite}
\usepackage{graphicx}

\newcommand{\cRB}{\color{RoyalBlue}}

\begin{document}

\vspace*{-2cm}
\centerline{\Large\textbf{The bulk composition of exo-planets}}
\bigskip
\noindent
Boris G\"ansicke (University of Warwick), John Debes (STScI), Patrick
Dufour (University of Montreal), Jay Farihi (UCL), Michael Jura
(UCLA), Mukremin Kilic (University of Oklahoma), Carl Melis (UCSD),
Dimitri Veras (University of Warwick), Siyi Xu (ESO), Ben Zuckerman
(UCLA)

\bigskip\noindent
{\large \cRB Motivation}

\smallskip\noindent Priorities in exo-planet research are rapidly
moving from finding planets to characterizing their physical
properties. Of key importance is their chemical composition, which
feeds back into our understanding of planet formation. Mass and radius
measurements of transiting planets yield bulk densities, from which
interior structures and compositions can be deduced
\citep{valenciaetal10-1}. However, those results are model-dependent
and subject to degeneracies \citep{rogers+seager10-1,
  dornetal15-1}. Transmission spectroscopy can provide insight into
the atmospheric compositions \citep{singetal13-1, demingetal13-1},
though cloud decks detected in a number of super earths systematically
limit the use of this method \citep{kreidbergetal14-1}.  \textit{For
  the foreseeable future, far-ultraviolet spectroscopy of white dwarfs
  accreting planetary debris remains the only way to directly and
  accurately measure the bulk abundances of exo-planetary bodies. The
  exploitation of this method is limited by the sensitivity of
  \textit{HST}, and significant progress will require a large-aperture
  space telescope with a high-throughput ultraviolet spectrograph. }

\medskip\noindent
{\large \cRB Evolved planetary systems}

\smallskip\noindent Practically all known planet host stars, including
the Sun, will evolve into white dwarfs, and many of their planets will
survive \citep{verasetal13-1, villaveretal14-1}. Observational
evidence for such evolved planetary systems includes the detection of
trace metals in the white dwarf photospheres \citep{koesteretal97-1},
and infrared and optical emission from circumstellar debris disks
\citep{zuckerman+becklin87-1, gaensickeetal06-3, farihietal09-1}. The
generally accepted model explaining these observation is the tidal
disruption of asteroids, minor planets, or planets \citep{jura03-1,
  debesetal12-1, verasetal14-1} perturbed onto star-crossing orbits by
dynamical interactions with planets \citep{debesetal02-1,
  frewen+hansen14-1, veras+gaensicke15-1}. Spectroscopic surveys now
unambiguously demonstrate that 25-50\% of white dwarfs host evolved
planetary systems \citep{zuckermanetal03-1, zuckermanetal10-1,
  koesteretal14-1}.

\medskip\noindent
{\large \cRB Debris-polluted white dwarfs as tracers of exo-planet
  bulk abundances}

\smallskip\noindent In a pioneering paper, \citet{zuckermanetal07-1}
showed that measuring the photospheric abundances of debris-polluted
white dwarfs provides an unrivaled window into the bulk composition
of exo-planetary material for planetary bodies with masses of
$10^{20}-10^{25}$\,g \citep{girvenetal12-1}, i.e. ranging from several
10\,km-sized asteroids to nearly the mass of Pluto. 

The ultraviolet wavelength range is fundamental for this work, as it
contains strong transitions of the rock-forming elements (Si, Fe, Mg,
O), refractory lithophiles (Ca, Al, Ti), and in particular of volatile
elements (C, N, P, S) that trace the formation region of the planetary
material relative to the snow line. We have led ten \textit{HST}/COS programs
that demonstrated the diagnostic potential of extra-solar
cosmochemistry using white dwarfs, corroborating the rocky,
volatile-depleted nature of the planetesimals \citep{juraetal12-1,
  xuetal13-1}, and detecting a variety in bulk compositions similar to,
if not exceeding, that seen among solar-system bodies \citep[][see
  Fig.\,1]{gaensickeetal12-1}. Noticeably, we have discovered
water-rich planetesimals \citep{farihietal13-2}, which provide the
potential for delivering water to planets in the habitable zone.

The measured planetary debris abundances provide important input into
our understanding of planet formation. Of particular importance for
the properties of planetary systems are the C/O and Mg/Si ratios. C/O
ratios $>0.8$ would result in a radically different setup from the
solar system, with O-chemistry replaced by C-chemistry, which is
discussed abundantly in the literature
\citep[e.g.][]{moriartyetal14-1}. The Mg/Si ratio determines the exact
composition of silicates, which in turn has implications for planetary
processes such as plate tectonics. Furthermore, the relative
abundances of Fe and siderophiles (Cr, Mn, S, Ni), and of refractory
lithophiles (Al, Ca, Ti) provides insight into the core and crust
formation, respectively \citep{jura+young14-1, melisetal11-1,
  dufouretal12-1}. In contrast to indirect measurements, such as
abundance studies of planet host stars
\citep[e.g.][]{delgado-menaetal10-1}, far-ultraviolet spectroscopy of
debris-polluted white dwarfs provides a \textit{direct measure} of
those ratios. The results from our published studies of have already
informed recent models of planet formation
\citep{carter-bondetal12-1}.

However, global insight into the chemistry of planetary systems will
only be possible from the detailed photospheric abundance studies of a
substantial number of white dwarfs. The current roster of planetary
debris abundance studies with at least five detected elements stands
at $\simeq15$ \citep{jura+young14-1}. With \textit{HST}, we may
double, maybe triple this number over the next couple of years, but
beyond that, the aperture of Hubble is too small to make significant
progress.

\begin{figure*}
\begin{minipage}[t]{8cm}
\includegraphics[width=8cm]{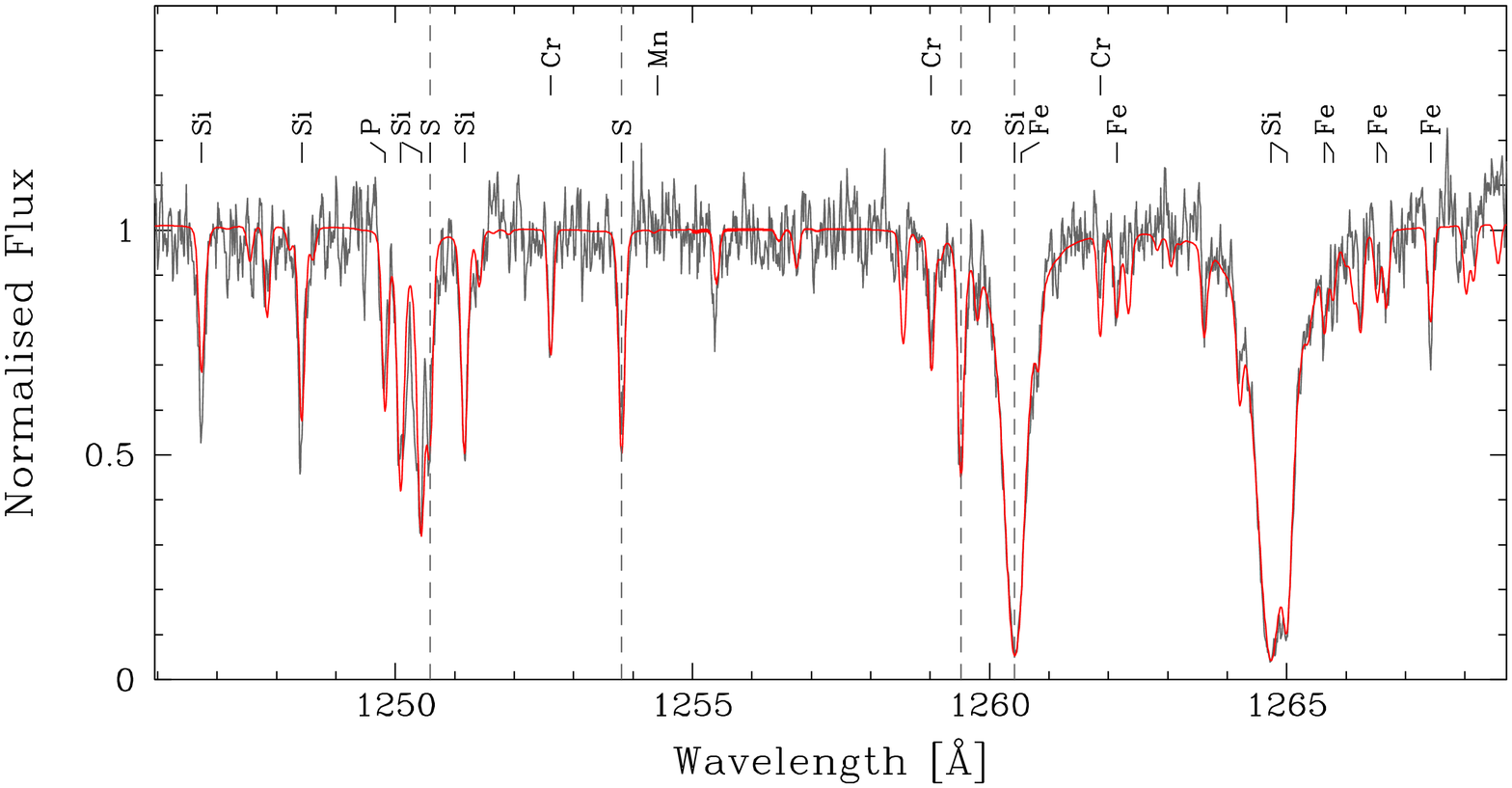}
\parbox{8cm}{\smallskip\small \textbf{Figure\,1}. Planetary debris is
  detected in far-ultraviolet spectroscopy of white dwarfs (top),
  providing bulk-abundances for exo-planetary bodies with masses of
  $\simeq10^{20}-10^{25}$\,g. About 15 systems studied so far are all
  ``rocky'', but show a large variety in their detailed compositions
  (right: large dots with error bars are extra-solar planetesimals,
  small dots are solar system meteorites).}
\end{minipage}
\begin{minipage}{8cm}
\includegraphics[width=8cm]{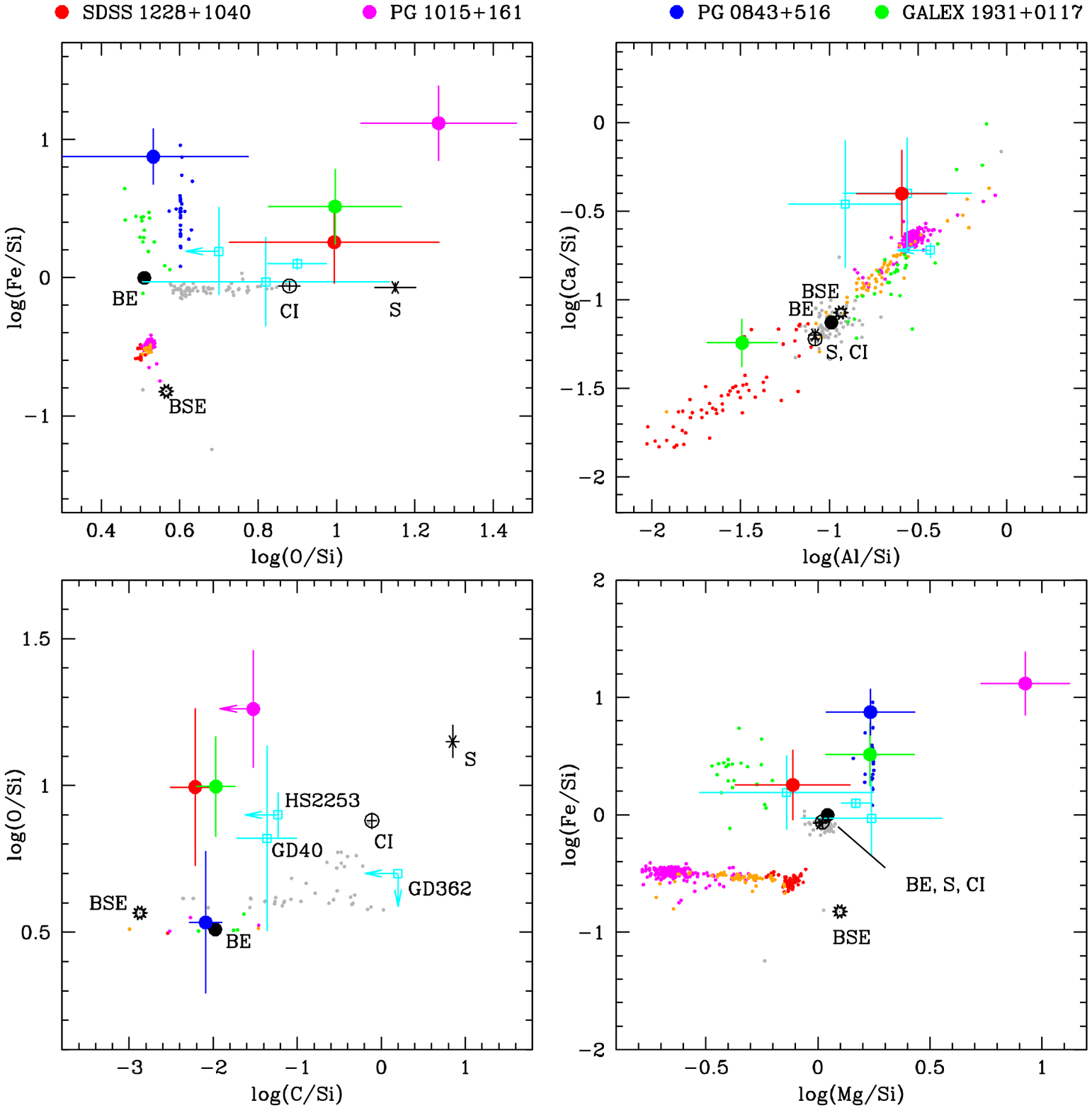}
\end{minipage}
\end{figure*}

\medskip\noindent
{\large \cRB Road-map for the next two decades, and the need for a
  large UV mission}

\smallskip\noindent Over the next four years, \textit{Gaia} will
identify $\simeq200\,000$ white dwarfs brighter than 20th
magnitude. This sample will be volume-limited out to 300pc for white
dwarfs with cooling ages of up to 600Myr, i.e. sufficiently hot to
have significant ultraviolet flux. Ground-based spectroscopic
follow-up of this sample (DESI/WEAVE/4MOST) will identify 1000s of
strongly polluted white dwarfs, but typically only provide abundances
for Ca and/or Mg. In addition, cross-correlation of the \textit{Gaia}
white dwarfs with the \textit{EUCLID} and \textit{WFIRST} surveys will
result in the detection of 100s, possibly 1000s of debris discs. The
detection of circumstellar debris is a direct proxy for
metal-pollution at ultraviolet wavelengths. A subset of these debris
discs will be bright enough to be followed-up with \textit{JWST}/MIRI,
providing detailed mineralogy. Those systems are particularly
valuable, as the dust mineralogy can be directly compared to atomic
abundances obtained from far-ultraviolet spectroscopy of the
debris-polluted white dwarfs.

In other words, the known sample of evolved planetary systems will
explode, similar to the dramatic increase in the number of planets
around main-sequence stars we witness thanks to missions like
\textit{Kepler}, \textit{TESS}, and \textit{PLATO}.  To fully exploit
the potential of evolved planetary systems for a detailed, large-scale
statistical study of the bulk abundances of exo-planetary systems
requires a large-aperture ultraviolet mission that can provide
follow-up spectroscopy for several hundred debris-polluted white
dwarfs.


\medskip\noindent
{\large \cRB Instrumental requirements}

\smallskip\noindent
Assuming a factor 30 increase in sensitivity compared to COS
($\times$15 for a 10\,m aperture, and $\times2$ from improved optics,
and improved orbital visibility) will increase the available volume
for detailed abundance studies by a factor $\simeq150$ compared to
what can be reached with \textit{HST}, sufficient to include $>1000$
potential targets for high-quality ultraviolet spectroscopic follow-up.

\begin{itemize}
\item Spectral resolution. A resolution of at least 20\,000, better
  50\,000 is necessary to resolve photospheric and interstellar
  features, and to avoid blending of lines. 
\item Spectral range. The "traditional" far-ultraviolet range 1100 to
  1800\,\AA\ contains most of the relevant atomic
  transitions. Extending coverage to 950\,\AA, i.e. including
  Ly$\beta$ and Ly$\gamma$, would greatly improve the atmospheric
  parameters, $T_\mathrm{eff}$ and $\log g$, which, in turn, result in
  more accurate diffusion velocities, and finally abundances. The
  shorter-wavelength range also contains a number of higher-ionization
  lines detected in hotter, younger white dwarfs.
\item Signal-to-noise ratio. The abundance measurements require a
  minimum S/N of 40 to model the strongest absorption lines. Detection
  of trace species not observed so far (e.g. N, rare earth elements)
  will need higher S/N, of up to 100, which is currently difficult to
  obtain with COS because of the limited telescope aperture, and
  fixed-noise patterns in the detectors.
\end{itemize}

\noindent
\bibliography{aamnem99,aabib,proceedings}
\bibliographystyle{aa}

\end{document}